\documentclass[aps,final,letterpaper,10pt,twocolumn,floats,showpacs,amsmath,amsfonts,amssymb]{revtex4-1}

\usepackage{graphicx}
\usepackage[colorlinks=true,citecolor=blue]{hyperref}
\usepackage[caption=false]{subfig}
\usepackage{epstopdf}
\usepackage{pstricks}
\usepackage{amsmath}
\usepackage{verbatim}
\begin{document}

\preprint{APS/123-QED}

\title{Coherence-enhanced constancy of a quantum thermoelectric generator}

\author{Krzysztof Ptaszy\'{n}ski}
 \email{krzysztof.ptaszynski@ifmpan.poznan.pl}
\affiliation{%
 Institute of Molecular Physics, Polish Academy of Sciences, ul. M. Smoluchowskiego 17, 60-179 Pozna\'{n}, Poland
}%

\date{\today}

\begin{abstract}
The study shows that presence of the quantum coherent, unitary component of the evolution of the system can improve constancy of heat engines, i.e., decrease fluctuations of the output power, in comparison with purely stochastic setups. This enables to overcome the recently derived trade-off between efficiency, power and constancy, which applies to classical Markovian steady-state heat engines. The concept is demonstrated using a model system consisting of two tunnel-coupled orbitals (i.e., electronic levels), each attached to a separate electronic reservoir; such a setup can be realized, for example, using quantum dots. Electronic transport is studied by means of the exact Levitov-Lesovik formula in the case without the Coulomb interaction between electrons, as well as applying a quantum master equation in the interacting case. Constancy of the analyzed thermoelectric generator is increased due to the fact that tunneling between the orbitals is associated with a unitary evolution of the electron state instead of a stochastic Poisson transition. This reduces stochasticity of the system, thus suppressing the current and power fluctuations. Moreover, noise can be further reduced by the Coulomb interaction between electrons which prevents the double occupancy of the system.

\end{abstract}

\maketitle


\section{\label{sec:intro}Introduction}
Heat engines realized in nanoscopic systems have received much attention in the last years due to both fundamental and practical reasons~\cite{benenti2017}. From a fundamental point of view, nanoscopic devices are intrinsically stochastic, and therefore fluctuations play an important role in their behavior~\cite{seifert2012, verley2014, esposito2015}. Of a special interest are universal laws governing the fluctuations, such as fluctuation theorems~\cite{seifert2012, verley2014, jarzynski1997, crooks1999, verley2014b, campsini2014} or bounds on statistical moments of thermodynamic currents~\cite{barato2015prl, pietzonka2016, gingrich2016, polettini2016, pietzonka2016b, pietzonka2017b, horowitz2017, pietzonka2017}. The other important issue is the role played by the quantum mechanical effects~\cite{kosloff2014}, which relates the topic to the emerging field of quantum thermodynamics~\cite{brandao2015, millen2016, vinjanampathy2016}. From an application-oriented perspective, nanodevices such as quantum dots are considered as promising candidates for efficient thermoelectric generators~\cite{sothmann2014, josefsson2017}.

Many previous studies have been concerned with the question whether the quantum effects can improve the performance of nanoscopic heat engines. In some cases the answer is positive: specific systems in which the quantum coherence can enhance the efficiency~\cite{scully2010, harbola2012, brandner2015, chen2017} or power~\cite{scully2011, harbola2012, hardal2017, bengtsson2018, klatzow2017} have been presented. On the other hand, there exist counterexamples of the thesis, in which the quantum mechanical effects are detrimental by reducing the efficiency~\cite{feldmann2003, karimi2016, chen2017} or power~\cite{karimi2016}, or increasing the noise~\cite{roulet2017}. Moreover, the reduction of power~\cite{brandner2017b} or efficiency~\cite{brandner2016} of cyclic heat engines operating in the linear response regime due to the quantum coherence has been shown to be universal. It seems therefore that the answer to the question posed depends on the specific physical situation and the quantity which one wants to optimize.

Here I present an example of the positive influence of the quantum coherence on performance of heat engines: it can increase their constancy (also referred to as a stability~\cite{holubec2014}), i.e., reduce fluctuations of the output power. This enables to overcome the trade-off between efficiency, power and constancy [Eq.~\eqref{tradeoff}] which has been derived by Pietzonka and Seifert~\cite{pietzonka2017} for classical, purely stochastic Markovian heat engines operating at the steady state. The concept is demonstrated using a model system of thermoelectric generator based on two tunnel-coupled orbitals, each attached to a separate electronic reservoir. Such a system can be realized, for example, using a double quantum dot molecule coupled in series to the leads or a single quantum dot attached to the spin-polarized electrodes. Statistical properties of the current flowing between the reservoirs are studied by means of the exact Levitov-Lesovik formula in the noninteracting case, as well as applying a quantum master equation, which enables to take into account the Coulomb interaction between electrons. Transport in the system is shown to be strongly influenced by presence of the coherent oscillations between the orbitals which, due to the unitary character of evolution of the electron state, reduce the current and power fluctuations. This increases the constancy of the thermoelectric generator in comparison with the classical case. The Coulomb interaction between electrons enables to further suppress the power fluctuations by preventing the double occupancy of the system.

The paper is organized as follows. In Sec.~\ref{sec:model} the analyzed model is described. Section~\ref{sec:nonint} discusses the case without the Coulomb interaction between electrons, for which the exacts results are obtained. In Sec.~\ref{sec:inter} I analyze the interacting case using a quantum master equation. Finally, Sec.~\ref{sec:conclusions} brings conclusions following from my results.

\section{\label{sec:model}Model}
I consider a spinless fermionic system consisting of two tunnel-coupled orbitals (i.e., electronic levels), each attached to a separate noninteracting semi-infinite reservoir; this will be referred to as the two-level bridge. Current fluctuations in such a system has been already studied in Ref.~\cite{esposito2015}, however without focusing on the phenomenon of noise suppression. The Hamiltonian of the analyzed model reads
\begin{align} \label{hamtotal} \nonumber
\hat{H}_b = & \sum_{\alpha} \epsilon_{\alpha} d_\alpha^\dagger d_\alpha +\Omega (d_L^\dagger d_R+d_R^\dagger d_L)+U d^\dagger_L d_L d^\dagger_R d_R \\
& +\sum_{\alpha \mathbf{k}} \epsilon_{\alpha \mathbf{k}} c_{\alpha \mathbf{k}}^{\dagger} c_{\alpha \mathbf{k}}+\sum_{\alpha \mathbf{k}} \left(t_{\alpha} c_{\alpha \mathbf{k}}^{\dagger} d_{\alpha} +t^*_{\alpha} d_{\alpha}^\dagger c_{\alpha \mathbf{k}} \right),
\end{align}
in which $d^\dagger_\alpha$ ($d_\alpha$) is the creation (annihilation) operator of the electron in the orbital $\alpha$ (with $\alpha \in \{L,R\}$), $\epsilon_{\alpha}$ is the energy of the orbital $\alpha$, $\Omega$ is the tunnel coupling between the orbitals (here, without loss of generality, taken to be a real number) and $U$ is the Coulomb interaction between the electrons occupying the orbitals L and R. The fourth term of the Hamiltonian describes electrons in the reservoirs; $\epsilon_{\alpha \mathbf{k}}$ denotes the energy of the electron in the lead $\alpha$ with a wave vector $\mathbf{k}$ and $c_{\alpha \mathbf{k}}^{\dagger}$ ($c_{\alpha \mathbf{k}}$) is the creation (annihilation) operator associated with such an electron. The last term describes tunneling between the lead $\alpha$ and the orbital $\alpha$, with $t_{\alpha}$ being the corresponding tunnel coupling. It is useful to define the coupling strength between the orbital $\alpha$ and the lead $\alpha$ as $\Gamma_{\alpha}=2 \pi |t_{\alpha}|^2 \rho_{\alpha}$, where $\rho_{\alpha}$ is the density of states in the lead $\alpha$. Here, for the sake of simplicity, I assume $\Gamma_\alpha$ to be energy-independent (the so-called wide-band limit).

%
\begin{figure} 
	\centering
	\subfloat[]{\includegraphics[width=0.8\linewidth]{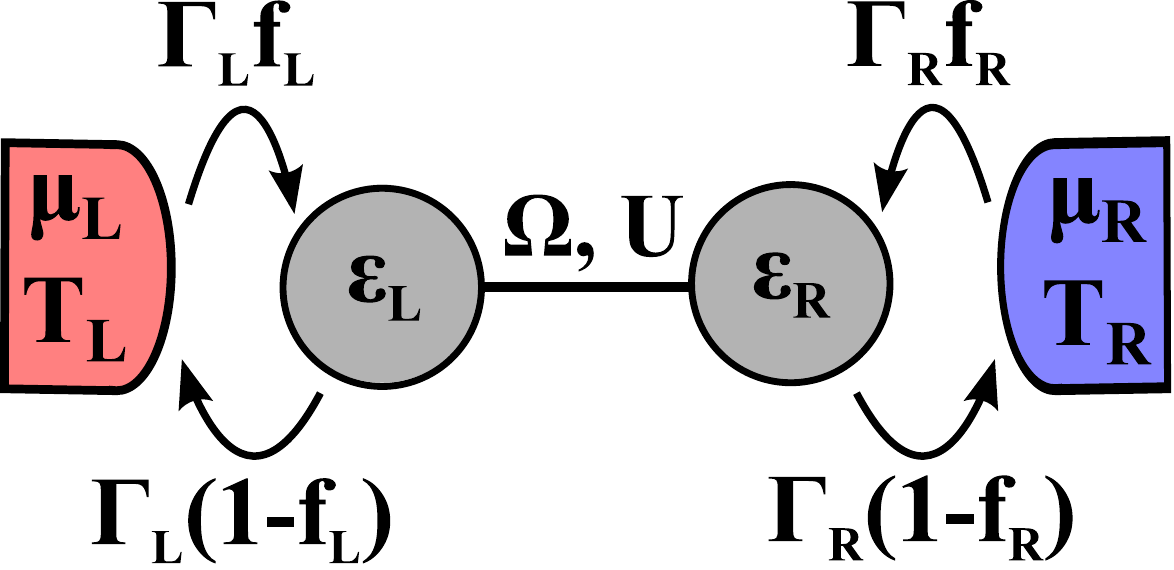}} \\
	\subfloat[]{\includegraphics[width=0.72\linewidth]{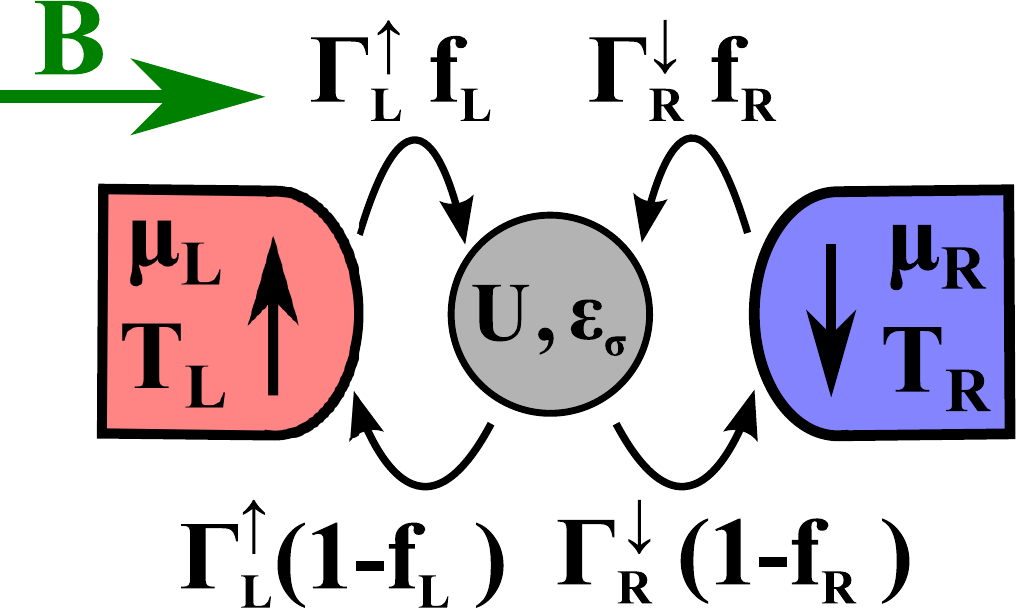}} 
	\caption{Different physical realizations of the two-level bridge described by the Hamiltonian~\eqref{hamtotal}. (a)~Double quantum dot molecule coupled in series to the leads; $f_\alpha={f[(\epsilon_\alpha-\mu_\alpha)/k_B T_\alpha]}$ denotes the Fermi distribution of electrons in the lead $\alpha$, other symbols explained below Eq.~\eqref{hamtotal}. (b)~Quantum dot attached to two fully spin-polarized leads with anti-parallel directions of magnetization, placed in the magnetic field perpendicular to the direction of magnetization.}
	\label{fig:uklad}
\end{figure}
%
The considered Hamiltonian is quite generic and may correspond to different physical systems. The most intuitive realization is a double quantum dot coupled in series to the leads [Fig.~\ref{fig:uklad}~(a)], which has been already thoroughly theoretically studied~\cite{gurvitz1996, gurvitz1998, wunsch2005, luo2011, hartle2014, seja2016, kirsanskas2018}; it can be made effectively spinless by applying a strong magnetic field, and thus removing one of the spin states out of the transport window. The other one is a spinfull single-level quantum dot attached to the fully spin-polarized leads with anti-parallel directions of magnetization, placed in the magnetic field oriented perpendicularly to the magnetization [Fig.~\ref{fig:uklad}~(b)]; the magnetic field induces the coherent oscillations between the spin states, thus playing a role of the tunnel coupling between the orbitals. Such a system can be easily mapped onto the considered model~\cite{wunsch2005} (cf. Appendix~\ref{sec:hamsv}). Similar quantum-dot-based spin valves have been a subject of many theoretical studies~\cite{rudzinski2001, konig2003, braun2004, braun2005, braun2006, sothmann2010, sothmann2014b, tang2018}. In particular, the recent paper of Stegmann \textit{et al.}~\cite{stegmann2018} have dealt with a problem similar to the considered in this article, i.e., the influence of the interplay of the stochastic electron tunneling and the coherent spin dynamics on the full counting statistics of the transmitted charge. A similar Hamiltonian has been also used to describe exciton transport through two coupled chromophores~\cite{goldberg2018}. One should be aware that the considered model neglects the environment-induced decoherence which may be important in real situations.

For the sake of comparison, I will also consider a single-level system described by the Hamiltonian
\begin{align} \label{hamtotalsing}
\hat{H}_s = \epsilon d^\dagger d  +\sum_{\alpha \mathbf{k}} \epsilon_{\alpha \mathbf{k}} c_{\alpha \mathbf{k}}^{\dagger} c_{\alpha \mathbf{k}}+\sum_{\alpha \mathbf{k}} \left(t_{\alpha} c_{\alpha \mathbf{k}}^{\dagger} d +t^*_{\alpha} d^\dagger c_{\alpha \mathbf{k}} \right),
\end{align}
where, in analogous way, $d^\dagger$ ($d$) is the creation (annihilation) operator of the electron in the orbital and $\epsilon$ is the orbital energy; $t_\alpha$ denotes then the coupling between the lead $\alpha$ and the orbital. Coupling strength $\Gamma_{\alpha}=2 \pi |t_{\alpha}|^2 \rho_{\alpha}$ is defined as in the two-level bridge.
 
\section{\label{sec:nonint}Noninteracting case} 
\subsection{\label{subsec:methodsnon}Methods}
I first focus on the system without the Coulomb interaction between electrons ($U=0$) for which exact results can be obtained; this will be referred to as the noninteracting case. Current fluctuations are analyzed using the full counting statistics formalism (see Ref.~\cite{esposito2009b} for the review). Let us denote the number of electrons tunneling from the left to the right lead in the time interval $[0,t]$ minus the number of electrons transported in the reverse direction as $n(t)$; the probability that $n(t)$ equals $N$ at the time $t$ is denoted as $P(N,t)$. The long time properties of the particle current fluctuations are described by the scaled cumulant generating function 
\begin{align}
\chi(\lambda)=\lim_{t \rightarrow \infty} \ln \left[\sum_{N=-\infty}^\infty P(N,t) e^{\lambda N} \right]/t,
\end{align}
which for noninteracting systems can be determined using the Levitov-Lesovik formula~\cite{levitov1993, levitov1996, levitov2004}
\begin{align} \label{levitov} \nonumber 
\chi(\lambda)=\int_{-\infty}^{\infty} \frac{d \omega}{2\pi} & \ln \left \{1+\mathcal{T}(\omega) \left [\left (e^{\lambda}-1 \right) f_L (\omega) g_R (\omega) \right. \right. \\ 
& \left. \left.  +\left (e^{-\lambda}-1 \right) f_R (\omega) g_L (\omega) \right] \right \},
\end{align}
where $\mathcal{T}(\omega)$ is the transmission function of the system, $f_\alpha (\omega)={f[(\omega-\mu_\alpha)/k_B T_\alpha]}$ is the Fermi distribution of electrons in the lead $\alpha$ (with $\mu_\alpha$ and $T_\alpha$ being the electrochemical potential and the temperature of the lead $\alpha$, respectively) and $g_\alpha (\omega)=1-f_\alpha (\omega)$. Here and from hereon I take $\hbar=1$. Equation~\eqref{levitov} has been shown to be exact in the situation in which each reservoir is tunnel-coupled to only one of the orbitals~\cite{esposito2015} (as in the considered case). Using this formula one may determine the scaled cumulants of the particle current defined as $c_i=\lim_{t \rightarrow \infty} C_i(t)/t$, where $C_i(t)$ is the $i$th cumulant of the distribution $P(N,t)$. They are expressed as
\begin{align} \label{cumcacl}
c_i = \left [ \frac{\partial^i \chi (\lambda)}{\partial \lambda^i} \right]_{\lambda=0}.
\end{align}
In particular, the mean particle current $\langle I \rangle=c_1=\lim_{t \rightarrow \infty} \langle n(t) \rangle/t$ and the particle current variance $c_2=\lim_{t \rightarrow \infty} \langle [\Delta n(t)]^2 \rangle/t$, where $\Delta n(t)=n(t)-\langle n(t) \rangle$, read as
\begin{align} \label{meancur}
\langle I \rangle  &=  \int_{-\infty}^{\infty} \frac{d \omega}{2 \pi} \mathcal{T}(\omega) [f_L (\omega)-f_R (\omega)], \\ \nonumber
c_2  &= \int_{-\infty}^{\infty} \frac{d \omega}{2 \pi} \mathcal{T}(\omega) \{ f_L (\omega)+f_R (\omega)-2f_L (\omega)f_R (\omega) \\ &-\mathcal{T}(\omega) [f_L (\omega)-f_R (\omega)]^2\}.
\end{align}
The scaled cumulant generating function for the heat flow to the lead $\alpha$ is obtained by multiplying the counting field $\lambda$ in the right-hand side of Eq.~\eqref{levitov} by $\Delta_\alpha$, where ${\Delta_L=\mu_L-\omega}$ and ${\Delta_R=\omega-\mu_R}$~\cite{pilgram2004, esposito2015}. In particular, the mean heat current to the lead $\alpha$ is given by the expression
\begin{align} \label{heatcur}
\langle \dot{Q}_\alpha \rangle = \int_{-\infty}^{\infty} \frac{d \omega}{2 \pi}\Delta_\alpha \mathcal{T}(\omega) [f_L (\omega)-f_R (\omega)].
\end{align}

Transmission function of the two-level bridge reads~\cite{sumetskii1991, zedler2009}
\begin{align} \label{transbr}
\mathcal{T}_{b}(\omega)=\frac{\Gamma_L \Gamma_R \Omega^2}{|(\omega-\epsilon_L+i \Gamma_L/2)(\omega-\epsilon_R+i \Gamma_R/2)-\Omega^2|^2}.
\end{align}
For comparison, transmission function of the single-level system is given by a well-known Breit-Wigner formula~\cite{nazarov2009}
\begin{align} \label{transsing}
\mathcal{T}_{s}(\omega)=\frac{\Gamma_L \Gamma_R}{(\omega-\epsilon)^2+(\Gamma_L+\Gamma_R)^2/4}.
\end{align}

In most cases, integrals in Eqs.~\eqref{meancur}--\eqref{heatcur} have to be evaluated in a numerical way. Analytic results can be obtained, however, in the weak coupling limit with $\Gamma, |\Omega| \ll k_B T_\alpha$, for the case of equal orbital energies $\epsilon_L=\epsilon_R=\epsilon$. In such a situation the Fermi distributions $f_\alpha(\omega)$ can be assumed to be constant in the range of $\omega$ for which the transmission is non-negligible, and therefore one can pull the functions $f_\alpha=f_\alpha(\epsilon)$ out of the integral. As a result, one obtains
\begin{align} \label{meancurweak}
\langle I \rangle & = \tilde{\mathcal{T}}_1 (f_L-f_R), \\ \label{c2weak}
c_2 & =\tilde{\mathcal{T}}_1 (f_L+f_R-2 f_L f_R) -\tilde{\mathcal{T}}_2 (f_L-f_R)^2, \\ \label{heatweak}
\langle \dot{Q}_\alpha \rangle & = \delta_\alpha \tilde{\mathcal{T}}_1 (f_L-f_R) = \delta_\alpha \langle I \rangle, 
\end{align}
where $\delta_L=\mu_L-\epsilon$, $\delta_R=\epsilon-\mu_R$ and
\begin{align}
\tilde{\mathcal{T}}_n=\int_{-\infty}^{\infty} \frac{d \omega}{2 \pi} \mathcal{T}(\omega)^n.
\end{align}
For the two-level bridge the parameters $\tilde{\mathcal{T}}_1$ and $\tilde{\mathcal{T}}_2$ are given by very complex expressions, expect for the symmetric case with  $\Gamma_L=\Gamma_R=\Gamma$, for which
\begin{align} \label{t1b}
\tilde{\mathcal{T}}_{1,b} &=\frac{2 \Gamma \Omega^2}{\Gamma^2+4 \Omega^2}, \\ \label{t2b}
\tilde{\mathcal{T}}_{2,b} &=\frac{4 \Gamma \Omega^4 (5 \Gamma^2+4 \Omega^2)}{(\Gamma^2+4 \Omega^2)^3}.	
\end{align}
For the single-level system, the parameters $\tilde{\mathcal{T}}_1$ and $\tilde{\mathcal{T}}_2$ take the simple form
\begin{align}
\tilde{\mathcal{T}}_{1,s} &=\frac{\Gamma_L \Gamma_R}{\Gamma_L+\Gamma_R}, \\
\tilde{\mathcal{T}}_{2,s} &=\frac{2\Gamma_L^2 \Gamma_R^2}{(\Gamma_L+\Gamma_R)^3}.	
\end{align}

\subsection{\label{subsec:resultsnon}Results}
Let us now analyze the results. For the sake of simplicity, in the whole section the symmetric coupling to the leads ($\Gamma_L=\Gamma_R=\Gamma$) and equal orbital energies ($\epsilon_L=\epsilon_R=0$) are assumed. I begin the analysis by showing that the quantum coherence in the two-level bridge enables to reduce the current fluctuations below the thermodynamic bound derived for classical, purely stochastic Markovian systems obeying the local detailed balance condition, according to which for an arbitrary fluctuating thermodynamic current $J_\nu$ (e.g., particle, charge or heat current) the following relation holds~\cite{horowitz2017}:
\begin{align} \label{uncert}
\frac{\text{Var}(J_\nu)}{\langle J_\nu \rangle^2}\geq \frac{2 k_B}{\langle \dot{s} \rangle},
\end{align}
where $\langle J_\nu \rangle$ is the mean current, $\langle \dot{s} \rangle$ is the mean rate of the entropy production in the whole system and $\text{Var}(J_\nu)={\langle \{\int_0^t [J_\nu (t')-\langle J_\nu \rangle] dt' \}^2 \rangle/ t}$, with $J_\nu (t')$ being the instantaneous current in the moment $t'$, is the normalized variance of the current integrated over the time interval $[0,t]$. This bound is valid for an arbitrary choice of the integration time $t$~\cite{horowitz2017}; however, for the sake of simplicity, from hereon I will focus on the long time limit with $t \rightarrow \infty$. For the considered system, the bound for fluctuations of the particle current takes the form
\begin{align} \label{boundpart}
\frac{c_2}{\langle I \rangle^2} \geq \frac{2 k_B}{\langle \dot{s} \rangle},
\end{align}
where $\langle I \rangle$ and $c_2$ are the mean particle current and the particle current variance, respectively, and the entropy production rate reads
\begin{align} \label{entrprod}
\langle \dot{s} \rangle = \frac{\langle \dot{Q}_L \rangle}{T_L}+\frac{\langle \dot{Q}_R \rangle}{T_R}.
\end{align}
As follows from Eqs.~\eqref{meancur} and~\eqref{heatcur}, for  equal temperatures of the left and the right lead ($T_L=T_R=T$) the entropy production is proportional to the mean particle current: ${\langle \dot{s} \rangle=(\langle \dot{Q}_L \rangle+\langle \dot{Q}_R \rangle)/T}={\langle I \rangle (\mu_L-\mu_R)/T}$. Inserting this formula into Eq.~\eqref{boundpart}, multiplying both sides by $\langle I \rangle$ and using a definition of the Fano factor
\begin{align} \label{fanodef}
F = \lim_{t \rightarrow \infty} \frac{\langle [\Delta n(t)]^2 \rangle}{|\langle n(t) \rangle|}=\frac{c_2}{|\langle I \rangle|},
\end{align}
the bound simplifies to the form
\begin{align} \label{boundsim}
F \geq \frac{2 k_B T}{|eV|},
\end{align}
where $eV=\mu_L-\mu_R$.

%
\begin{figure} 
	\centering
	\includegraphics[width=0.9\linewidth]{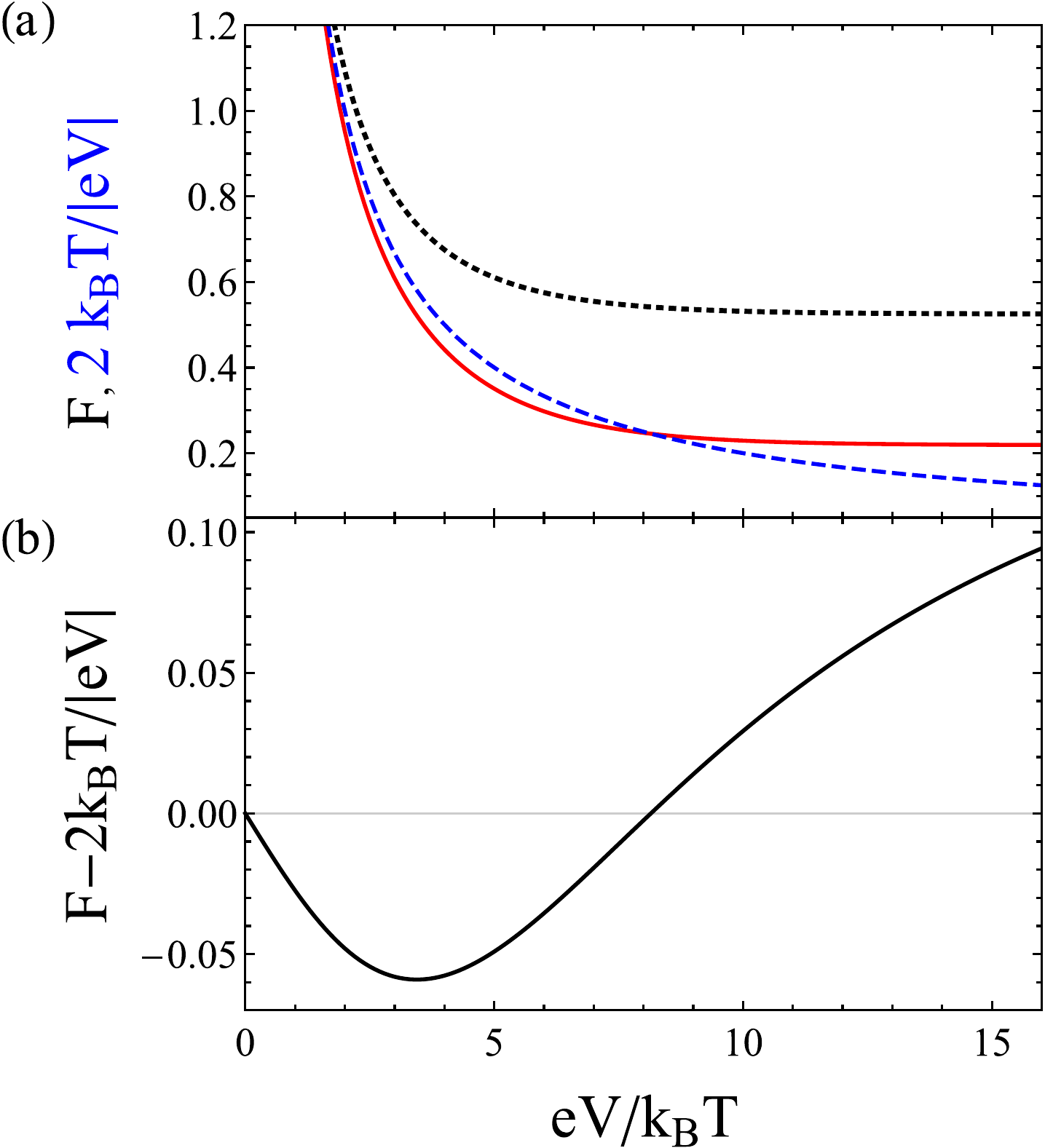} 
	\caption{(a) Fano factor $F$ for the two-level bridge (red solid line) and the single level (black dotted line) as a function of the voltage $V$ compared with the classical bound $2k_B T/|eV|$ [cf. Eq.~\eqref{boundsim}, blue dashed line] for $\mu_L=-\mu_R=eV/2$, $T_L=T_R=T$, $\Gamma=0.01 k_B T$ and $|\Omega|=\sqrt{15}\Gamma/6 \approx 0.65 \Gamma$ (i.e., for the value of $|\Omega|$ for which $F$ is minimized). (b) Difference of the Fano factor and the classical bound for the two-level bridge for the same values of parameters.}
	\label{fig:fano}
\end{figure}
%
Let us now consider the conditions for which the Fano factor in the two-level bridge is minimized, focusing on the weak coupling regime. Equations~\eqref{meancurweak}--\eqref{c2weak} and~\eqref{fanodef} indicate that this occurs when the ratio $\tilde{\mathcal{T}}_2/\tilde{\mathcal{T}}_1$ is maximized. As follows from Eqs.~\eqref{t1b}--\eqref{t2b}, this takes place for $|\Omega|=\sqrt{15} \Gamma/6 \approx 0.65 \Gamma$, for which the timescale of the coherent oscillations between the orbitals is comparable to the timescale of the lead-orbital tunneling. Figure~\ref{fig:fano} shows the value of the Fano factor for the two-level bridge and the single-level system compared with the classical bound as a function of the voltage $V$ for $T_L=T_R=T$, relatively small $\Gamma=0.01 k_B T$ and the optimal value of $|\Omega|=\sqrt{15} \Gamma/6$. It can be clearly seen that in the certain range of the voltage ($eV \lessapprox 8 k_B T$) the Fano factor for the two-level bridge (red solid line) is reduced below the classical bound given by Eq.~\eqref{boundsim} (blue dashed line). On the other hand, this is not observed for the single-level system (black dotted line). 

The mechanism of the noise suppression in the two-level bridge can be explained as follows: Since the tunnel coupling to the leads is relatively weak, one observes a sequential electron tunneling between the electrodes and the orbitals~\cite{buttiker1986}, which can be described as a stochastic Markovian process~\cite{averin1991, li2005}. For the single-level system transport of the electron between the reservoirs consists of two steps: (1)~sequential tunneling from the left lead to the orbital, (2)~sequential tunneling from the orbital to the right lead. For such two-step Markovian processes the Fano factor obeys the well known bound $F \geq 1/2$~\cite{hershfield1993, korotkov1994, blanter2000}. On the other hand, when one neglects for a while the double occupancy of the system, transfer of the electron through the two-level bridge may be described as consisting of three steps: (1)~sequential tunneling of the electron from the left lead to the empty left orbital, (2)~coherent tunneling from the occupied left to the empty right orbital, (3)~sequential tunneling from the right orbital to the right lead (taking into account that due to the Pauli principle only tunneling to the empty orbitals is possible). It is already known that by increasing the number of steps in the transport cycle one can reduce the current fluctuations~\cite{barato2015prl, koza2002, novotny2004, bulka2008, haupt2008}. However, increase of the number of steps does not fully explain the observed noise suppression, since for classical Markovian three-step processes the bound $F \geq 1/3$ holds~\cite{bulka2008}, which is not satisfied for the two-level bridge. Moreover, this does not explain the violation of the thermodynamic uncertainty relation [Eq.~\eqref{uncert}], which is valid for arbitrary classical Markovian system, independently of the number of steps. The decisive factor is the fact that tunneling between the orbitals is not associated with a stochastic Poisson transition, but with a coherent evolution of the electron state. This reduces stochasticity of the system, thus suppressing the noise. 

%
\begin{figure} 
	\centering
	\includegraphics[width=0.9\linewidth]{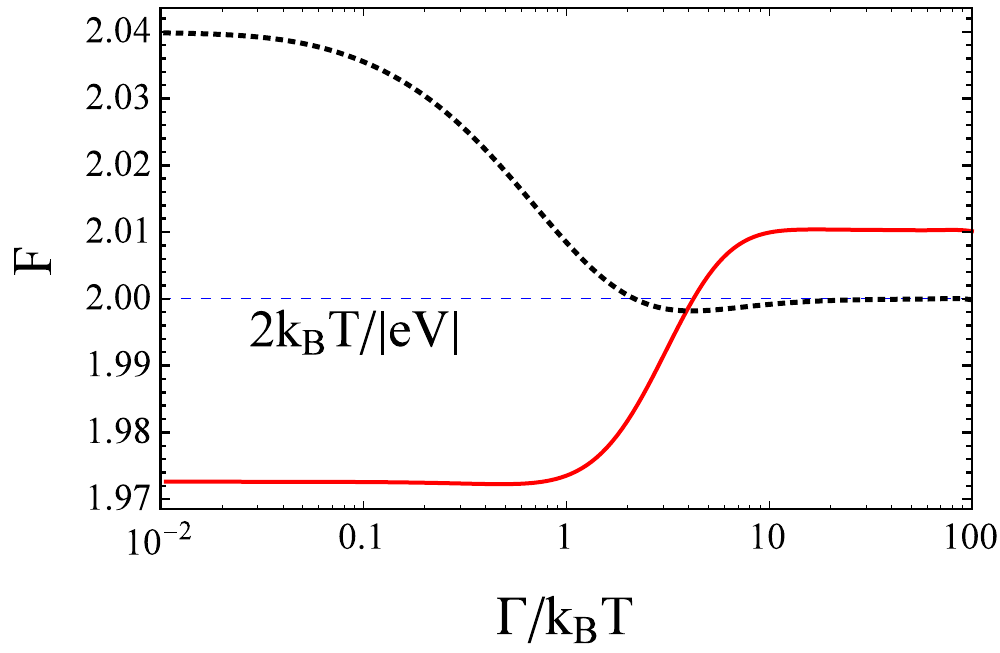} 
	\caption{Fano factor as a function of $\Gamma$ for the two-level bridge (red solid line) and the single-level system (black dotted line) compared to the classical bound $2 k_B T/|eV|$ (blue dashed line at $F=2$) for $T_L=T_R=T$, $\mu_L=-\mu_R=k_B T/2$ and $|\Omega|=\sqrt{15}\Gamma/6$.}
	\label{fig:fvsgamma}
\end{figure}
%
Next, I analyze how the current fluctuations are affected by the lead-orbital coupling $\Gamma$. As previously, $|\Omega|=\sqrt{15}\Gamma/6$ is taken. Figure~\ref{fig:fvsgamma} shows that in the two-level bridge the Fano factor is nearly constant up to $\Gamma \approx 0.1 k_B T$; in this range $F$ converges to the result provided by the weak coupling approximation [Eqs.~\eqref{meancurweak}--\eqref{heatweak}]. In the range $0.1k_B T \lessapprox \Gamma \lessapprox k_B T$ one observes a slight reduction of $F$, with a minimum for $\Gamma \approx k_B T$, which is due to the coherent nature of the lead-orbital tunneling; this effect is, however, barely observable. On the other hand, for large enough $\Gamma \gtrapprox k_B T$ noise is significantly enhanced, which may be ascribed to the increased level splitting caused by the tunnel coupling $\Omega$ (which is taken to be proportional to $\Gamma$). In comparison, in the single-level system for high values of $\Gamma$ the coherent nature of the lead-orbital tunneling suppresses the current fluctuations, enabling to overcome the classical bound for $\Gamma \approx 5 k_B T$. However, the reduction of the Fano factor is much less pronounced than in the two-level bridge with small $\Gamma$, although certainly higher than the numerical error; moreover, it is not observed for higher voltages $eV \equiv \mu_L-\mu_R \gtrapprox 4k_B T$ (not shown).

The fact, that the thermodynamic uncertainty relation does not hold beyond its range of validity, is in itself not surprising. The similar situation has been already observed for multi-terminal ballistic junctions~\cite{brandner2017} or the underdamped Brownian particles, for which suppression of fluctuations is the result of the classical, unitary Hamiltonian evolution~\cite{fischer2018}. In particular, violation of thermodynamic uncertainty relation in similar quantum coherent systems has been simultaneously demonstrated by an independent study of Agarwalla and Segal~\cite{agarwalla2018}, which also provided general conditions of this occurrence in terms of nonlinear transport coefficients. However, to the best of my knowledge, practical implications of this fact for the design of heat engines has been not yet studied. Here I will show that the noise suppression due to the quantum coherence enables to suppress the power fluctuations of nanoscopic thermoelectric generators, enabling to overcome the thermodynamic trade-off between efficiency, power and constancy derived by Pietzonka and Seifert~\cite{pietzonka2017} for classical, purely stochastic Markovian heat engines operating at the steady state, which implies that power fluctuations of the heat engine can be reduced only at the cost of reducing its efficiency. This bound reads
\begin{align} \label{tradeoff}
\frac{\text{Var}(P)}{\langle P \rangle} \geq k_B T_C \frac{2 \eta}{\eta_C-\eta},
\end{align}
where $T_C$ is the temperature of the cold reservoir, $\langle P \rangle$ and $\eta$ are the mean power and the efficiency of the heat engine, respectively, $\eta_C=1-T_C/T_H$ is the Carnot efficiency, where $T_H$ is the temperature of the hot reservoir, and $\text{Var}(P)$ is the power variance defined as
\begin{align}
\text{Var}(P) &= \lim_{t \rightarrow \infty} \left\langle \left\{\int_0^t [P(t')-\langle P \rangle] dt' \right\}^2 \right\rangle/ t \\ \nonumber
& = \lim_{t \rightarrow \infty} \langle [W(t)-\langle P \rangle t]^2 \rangle/t,
\end{align}
where $P(t')$ is the instantaneous power in the moment $t'$ and $W(t)=\int_0^t P(t') dt$ is the work done in the time interval $[0,t]$ (as mentioned, I focus on the long time limit with $t \rightarrow \infty$). This relation can be easily derived using Eq.~\eqref{uncert} by taking $J_\nu=P$ [since Eq.~\eqref{uncert} applies to an arbitrary current, including power] and inserting $\langle \dot{s} \rangle=\langle \dot{Q}_C \rangle/T_C+\langle \dot{Q}_H \rangle/T_H$, where $\langle \dot{Q}_C \rangle$ ($\langle \dot{Q}_H \rangle$) is the heat current flowing to the cold (hot) reservoir; by taking $\langle \dot{Q}_H \rangle=-\langle P \rangle/\eta$, $\langle \dot{Q}_C \rangle=-\langle \dot{Q}_H \rangle-\langle P \rangle={\langle P \rangle (1-\eta)/\eta}$ and $T_H={T_C/(1-\eta_C)}$ one obtains Eq.~\eqref{tradeoff}~\cite{pietzonka2017}.

To show, that Eq.~\eqref{tradeoff} is not valid any longer when the quantum coherence is present, I will consider the regime in which the two-level bridge acts as a thermoelectric generator, i.e., current is driven against the voltage due to the temperature gradient. Without loss of generality, let us focus on the case when ${T_L>T_R}$ and therefore ${T_C=T_R}$, ${T_H=T_L}$. Power of the generator is then equal to $\langle P \rangle= \langle I {\rangle (\mu_R -\mu_L)}$ and efficiency equals $\eta=-\langle P \rangle/\langle \dot{Q}_L\rangle$. The power variance can be calculated as
\begin{align}
\text{Var}(P) &=(\mu_R-\mu_L)^2 \lim_{t \rightarrow \infty} \langle [n(t)-\langle I \rangle t]^2 \rangle/t \\ \nonumber
&= (\mu_R-\mu_L)^2 c_2.
\end{align}
In particular, in the weak coupling regime (${\Gamma, |\Omega| \ll k_B T_\alpha}$) the system acts as a thermoelectric generator when $T_R/T_L < \mu_R/\mu_L < 1$, the mean heat current to the left lead reads $\langle \dot{Q}_L\rangle=\mu_L \langle I \rangle$ [cf. Eq.~\eqref{heatweak}] and efficiency equals $\eta={1-\mu_R/\mu_L}$; for $\mu_R/\mu_L \rightarrow T_R/T_L$ efficiency reaches the Carnot limit $\eta_C={1-T_R/T_L}$.

Let us now define the ratio of the right-hand side and the left-hand side of Eq.~\eqref{tradeoff} as the normalized constancy
\begin{align} \label{const}
C_N=k_B T_C \frac{\langle P \rangle}{\text{Var}(P)} \frac{2 \eta}{\eta_C-\eta},
\end{align}
which for the considered system it is equal to
\begin{align} \label{constsys}
	C_N=\frac{2 k_B T_R \langle I  \rangle^2}{c_2[\langle \dot{Q}_L \rangle (T_R/T_L-1)+(\mu_L-\mu_R) \langle I  \rangle]}.
\end{align}
In the weak coupling limit Eq.~\eqref{constsys} simplifies to the form
\begin{align} \label{constsysweak}
C_N=\frac{2 k_B \langle I \rangle}{c_2(\mu_L/T_L-\mu_R/T_R)}.
\end{align}
%
\begin{figure} 
	\centering
	\includegraphics[width=0.9\linewidth]{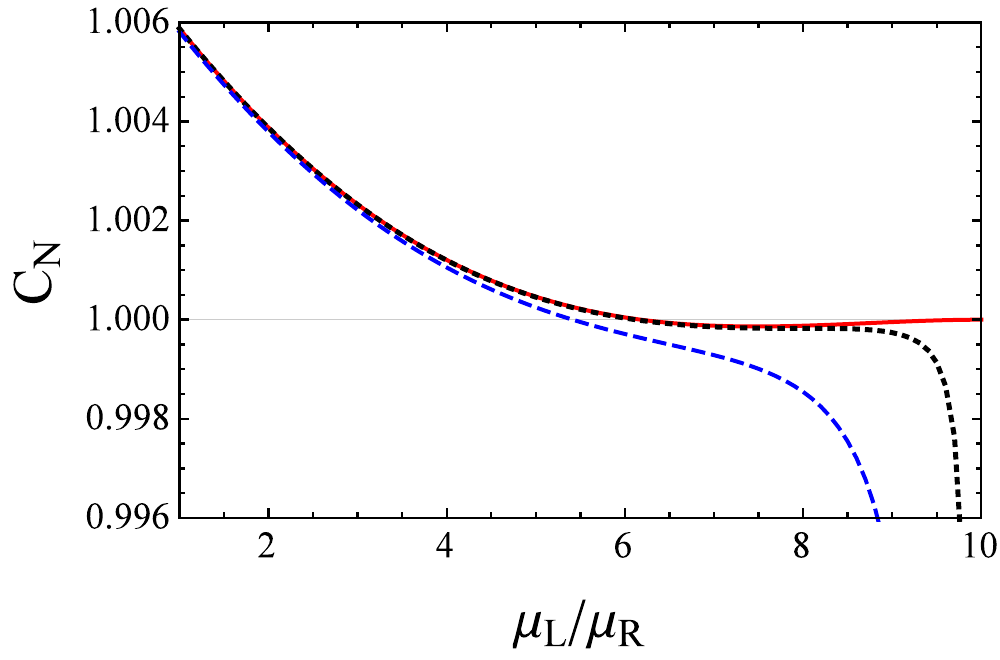} 
	\caption{Normalized constancy $C_N$ as a function of the ratio $\mu_L/\mu_R$ for the two-level bridge with $T_L=10 T_R$ calculated analytically in the limit of $\Gamma \rightarrow 0$ (red solid line) and numerically for $\Gamma=0.002 k_B T_R$ (black dotted line) and $\Gamma=0.01 k_B T_R$ (blue dashed line). All results for $\mu_R=k_B T_R$ and $|\Omega|=\sqrt{15} \Gamma/6$.}
	\label{fig:normconstnon}
\end{figure}
%
As follows from Eq.~\eqref{tradeoff}, in classical Markovian systems $C_N \leq 1$. Figure~\ref{fig:normconstnon} shows the normalized constancy of the two-level bridge for the case of relatively large temperature bias ($T_L=10 T_R$) calculated analytically in the weak coupling limit [using Eqs.~\eqref{meancurweak}--\eqref{heatweak}], as well as numerically for two small but finite values of the tunnel coupling ($\Gamma=0.002 k_B T_R$ and $\Gamma=0.01 k_B T_R$). In the case of small voltage bias (i.e., for the ratio $\mu_L/\mu_R$ relatively small in comparison with $T_L/T_R$), the normalized constancy is increased above the classical bound (i.e., $C_N$ exceeds 1). In this regime the analytic formula agrees well with the numerical results for both values of the lead-orbital coupling, which confirms that the result is not a consequence of the used approximation or the numerical error. On the other hand, when the ratio $\mu_L/\mu_R$ becomes comparable to $T_L/T_R$ (i.e., the thermal bias is counter-balanced by the voltage bias, such that current is suppressed) the analytic formula and the numerical results diverge; the deviation is larger for higher value of $\Gamma$. This can be explained as follows: For $\mu_L/\mu_R \approx T_L/T_R$ the Fermi distributions of the left and the right lead [denoted as $f_L (\omega)$ and $f_R(\omega)$, respectively] have similar values. In such a situation very subtle differences of $f_L (\omega)$ and $f_R(\omega)$ start to play an important role, and therefore the assumption that $f_\alpha(\omega)$ is constant in the vicinity of resonance, which was used to derive Eqs.~\eqref{meancurweak}--\eqref{heatweak}, is not valid any longer. The deviation is larger for higher value of $\Gamma$ due to the increased broadening of the transmission function.

Unfortunately, the increase of constancy above the classical limit is very small (below 1\%). Nevertheless, the results constitute the proof of concept: they show that the quantum coherence indeed enables to overcome the classical trade-off between efficiency, power and constancy. This motivates the quest for further optimization. In the next section I will show that constancy can be improved more significantly when the Coulomb interaction between electrons is present.

\section{\label{sec:inter}Interacting case} 

\subsection{\label{sec:methodsint}Methods}
Now I consider the system with the Coulomb interaction between electrons ($U \neq 0$); this will be referred to as the interacting case. The study focuses on the situation in which the coupling to the leads is weak, i.e., $\Gamma_\alpha \ll k_B T_\alpha$, such that one observes a Markovian sequential tunneling between the orbitals and the electrodes~\cite{averin1991, li2005}. It will be also assumed that $|\Omega| \approx \Gamma_{\alpha} \ll k_B T_\alpha$ and $|\epsilon_L-\epsilon_R| \ll k_B T_\alpha$, such that the timescale of the coherent oscillations between the orbitals is comparable to the timescale of the lead-orbital tunneling, whereas energy separation between the single-electron states is smeared by the temperature.
	
Transport in Coulomb-interacting systems is usually studied by means of a quantum master equation. However, most common master equation approaches to electronic transport either properly treat the coherences between states of the system, but assume the infinite bias regime in which the thermally excited tunneling against the bias is neglected (like the method of Gurvitz and Prager~\cite{gurvitz1996, gurvitz1998}), or work for finite bias, dealing with the energy-dependence of tunneling, but neglect the coherences (like the Pauli master equation~\cite{breuer2002}, also referred to as the diagonalized master equation~\cite{poltl2009}). Here I need an approach which both describes the coherent oscillations between the orbitals and works for finite bias. There are several more or less sophisticated methods to deal with such a case (see, eg., Refs.~\cite{benenti2009, busl2010, seja2016, kirsanskas2018, mitchison2018}). In this paper I apply the master equation derived by Wunsch \textit{et al.}~\cite{wunsch2005} using the real-time diagrammatic technique, which is particularly suitable for the parameter range considered in this paper. This approach is perturbative to the first order of the tunnel coupling $\Gamma_L+\Gamma_R$ and the zeroth order of the level splitting $\sqrt{\Omega^2+(\epsilon_L-\epsilon_R)^2}$. While the original paper of Wunsch \textit{et al.}~\cite{wunsch2005} focused on the analysis of the mean current at the stationary state, the later publication of Braun \textit{et al.}~\cite{braun2006} generalized this method to the calculation of the finite-frequency noise, whereas Luo \textit{et al.}~\cite{luo2011} analyzed the full counting statistics by deriving the equivalent number-resolved master equation in an alternative way. Details of the derivation of the master equation are well described in Refs.~\cite{wunsch2005, braun2006, luo2011}, and in particular in Secs. II--III and Appendixes A--B in Ref.~\cite{wunsch2005}. In Refs.~\cite{wunsch2005, braun2006, luo2011} the obtained master equation has been written in the form of equations of motions for the density matrix elements; here it is rewritten in the commonly used Lindblad form:
\begin{align} \label{lindblad}
& \frac{d \hat{\rho}}{dt} = -i \left[ \hat{H}_{c}, \hat{\rho} \right] \\ \nonumber  &+ \sum_{\alpha i} \left(L^{\dagger}_{\alpha i} \hat{\rho} L_{\alpha i}-L_{\alpha i} L^{\dagger}_{\alpha i} \hat{\rho}/2 -\hat{\rho} L_{\alpha i } L^{\dagger}_{\alpha i}/2 \right),
\end{align}	
where the density matrix of the system $\hat{\rho}$ is written in the basis of the localized states, i.e., ${\{|0\rangle, |L \rangle, |R \rangle , |D \rangle \}}$, in which $|L\rangle \equiv d^\dagger_L |0 \rangle$ and $|R\rangle \equiv d^\dagger_R |0 \rangle$ are the singly occupied states, whereas $|D \rangle \equiv d^\dagger_R d^\dagger_L |0 \rangle$ is the doubly occupied state. One can check the equivalence of the aforementioned equations of motions presented in Refs.~\cite{wunsch2005, braun2006, luo2011} with the ones produced by Eq.~\eqref{lindblad}.

The right-hand side of Eq.~\eqref{lindblad} consists of two parts. The first term describes the unitary evolution of the density matrix associated with the coherent oscillations between the orbitals. Here \begin{align} \label{hamc}
	\hat{H}_{c} = & \sum_{\alpha} \tilde{\epsilon}_{\alpha} d_\alpha^\dagger d_\alpha +\Omega (d_L^\dagger d_R+d_R^\dagger d_L)+U d^\dagger_L d_L d^\dagger_R d_R,
	\end{align}
is the effective Hamiltonian of the central region in which $\tilde{\epsilon}_{\alpha}$ are the renormalized orbital energies~\cite{wunsch2005}
\begin{align}
\tilde{\epsilon}_{\alpha}=\epsilon_\alpha+\phi_\alpha (\epsilon_m)-\phi_\alpha (\epsilon_m+U),
\end{align}
where $\epsilon_m=(\epsilon_L+\epsilon_R)/2$ is the mean orbital energy and
\begin{align}
\phi_\alpha (\omega)=\frac{\Gamma_{\alpha}}{2 \pi} \text{Re} \left[\Psi \left(\frac{1}{2} +i \frac{\omega-\mu_\alpha}{2 \pi k_B T_\alpha }\right) \right],
\end{align}
with $\Psi$ being the digamma function. The level renormalization, corresponding to the already thoroughly studied exchange coupling to the electrodes in quantum dot spin valves~\cite{konig2003, braun2004, braun2005, braun2006, sothmann2010, sothmann2014b, stegmann2018}, is a result of the interplay of the electron tunneling and the inter-orbital Coulomb interaction~\cite{wunsch2005, luo2011, hartle2014} (here I consider a spinless system, in which a single orbital can be at most singly occupied, and therefore the intra-orbital Coulomb interaction is not present); it vanishes for $U=0$. For a strongly interacting case ($U \gtrapprox k_B T_\alpha$) the magnitude of the level renormalization is comparable to $|\Omega|$ (due to $|\Omega| \approx \Gamma_{\alpha}$), and therefore the effect may strongly influence the coherent oscillations between the orbitals~\cite{konig2003, braun2004, braun2005, wunsch2005, braun2006, luo2011, sothmann2010, sothmann2014b, stegmann2018}.

The second term of Eq.~\eqref{lindblad} describes the sequential tunneling between the orbitals and the leads, which corresponds to the classical Markovian dynamics. Here, the Lindblad operators $L_{\alpha i}^\dagger$, $L_{\alpha i}$, with $\alpha \in \{L, R \}$ and $i \in \{1,2,3,4\}$, are defined as
\begin{subequations} \label{lindbladop}
\begin{align}
L_{\alpha 1}^\dagger &=\sqrt{\Gamma_\alpha f_\alpha} |\alpha \rangle \langle \alpha| c^\dagger_\alpha, \\
L_{\alpha 2}^\dagger &=\sqrt{\Gamma_\alpha f_\alpha^U} |D \rangle \langle D| c^\dagger_\alpha, \\
L_{\alpha 3}^\dagger &=\sqrt{\Gamma_\alpha (1-f_\alpha)} c_\alpha |\alpha \rangle \langle \alpha|, \\
L_{\alpha 4}^\dagger &=\sqrt{\Gamma_\alpha (1-f_\alpha^U)} c_\alpha |D \rangle \langle D|,
\end{align}
\end{subequations}
where $f_\alpha={f[(\epsilon_m-\mu_\alpha)/k_B T_\alpha]}$ and $f_\alpha^U={f[(\epsilon_m-\mu_\alpha+U)/k_B T_\alpha]}$. Operators $L_{\alpha 1}^\dagger$ and $L_{\alpha 2}^\dagger$ describe tunneling to the orbital $\alpha$ when the other orbital is either empty or occupied, respectively; the operators $L_{\alpha 3}^\dagger$ and $L_{\alpha 4}^\dagger$ correspond to the reverse processes. Here the tunneling rates are assumed to be dependent on the mean orbital energy $\epsilon_m$ rather than the energies of single orbitals or the eigenstate energies; this assumption is justified when $|\Omega|, |\epsilon_L-\epsilon_R|, \Gamma_\alpha \ll k_B T_\alpha$, and therefore the separation between the energies of single-particle states is smeared by the temperature~\cite{wunsch2005, luo2011}. Since the level renormalization is small in comparison with the temperature (i.e., $|\tilde{\epsilon}_\alpha-\epsilon_\alpha| \ll k_B T_\alpha$), I will later substitute $\epsilon_m \rightarrow (\tilde{\epsilon}_L+\tilde{\epsilon}_R)/2$, and then treat the renormalized orbital energies $\tilde{\epsilon}_\alpha$ as independent variables.

The used method, which operates in the basis of localized states and assumes that the tunnel coupling between the orbitals does not influence the lead-orbital tunneling, is commonly referred to as the local approach~\cite{levy2014, stockburger2016, trushechkin2016, hofer2017}. The question may arise whether such a method is legitimate. It has been previously shown that the local approach, although well justified in many cases~\cite{trushechkin2016, hofer2017}, may sometimes provide unphysical results~\cite{stockburger2016}. For example, it can violate the second law of thermodynamics, sometimes even in the regime of weak coherent coupling between states which is assumed in this section~\cite{levy2014}. It also does not provide the Gibbs-Boltzmann distribution at equilibrium~\cite{carmichael1973}. Moreover, as already mentioned, the applied approach is perturbative. It has been demonstrated that perturbative master equations approaches may exhibit thermodynamic inconsistencies, such as violation of the Onsager theorem~\cite{seja2016}.
	
A partial answer to the question about the validity of the applied approach may be provided by comparing the result given by the master equation with the exacts ones for the noninteracting case (as in Refs.~\cite{seja2016, kirsanskas2018}). Here, one can easily check that for the two-level bridge with $U=0$ and $\epsilon_L=\epsilon_R$ the analytic expressions for the steady-state quantities (such as current or noise) obtained using Eq.~\eqref{lindblad} are equivalent to those calculated applying the Levitov-Lesovik formula in the weak coupling limit [Eqs.~\eqref{meancurweak}--\eqref{heatweak}]. As shown in Figs.~\ref{fig:fvsgamma}, \ref{fig:normconstnon}, \ref{fig:constint} and~\ref{fig:powconst}~(b), this limit is well applicable as long as $\Gamma_{\alpha} \ll k_B T_\alpha $ and the difference of the Fermi distributions of the left and the right lead in the vicinity of resonance is sufficiently large (i.e., the system is sufficiently far from equilibrium). The noise suppression below the classical bound can be observed in the regime, in which all these conditions are met. This supports my claim that the applied perturbative master equation, although should be used with caution, may provide physically relevant results (in its range of applicability). The direct assessment of the validity of this approach for the interacting case would require the use of advanced numerically exact methods, such as the hierarchical quantum master equation~\cite{hartle2014} or the quantum Monte Carlo~\cite{ridley2018}, which goes beyond the scope of this paper.

Equation~\eqref{lindblad} can be rewritten in the Liouville space form~\cite{carmichael1993, breuer2002}
\begin{align} \label{mastereq}
\dot{\rho}(t)=\mathcal{W} \rho(t),
\end{align}
where $\rho(t)$ is the column vector containing both the diagonal and the non-diagonal elements of the density matrix $\hat{\rho}$ (the state probabilities and the coherences) and $\mathcal{W}$ is the square matrix representing the Liouvillian. Here the vector $\rho(t)$ is defined as $\rho(t)=(\rho_{00}, \rho_{LL}, \rho_{RR}, \rho_{DD}, \mathcal{R}_{LR}, \mathcal{I}_{LR})^T$, where $\rho_{ij}$ = $\langle i|\hat{\rho}|j \rangle$, $\mathcal{R}_{LR}=\text{Re}(\rho_{LR})$ and $\mathcal{I}_{LR}=\text{Im}(\rho_{LR})$; the evolution of the other elements of the density matrix is decoupled from the dynamics of populations, and therefore is neglected. Full matrix form of the Liouvillian $\mathcal{W}$ is presented in the Appendix~\ref{sec:matrform}.

The full counting statistics formalism for systems described by a master equation has been developed by Bagrets and Nazarov~\cite{bagrets2003}. Here I apply a recent version of this approach which enables to determine the scaled cumulants of the current using the characteristic polynomial of the counting-field-dependent Liouvillian of the system instead of the scaled cumulant generating function~\cite{bruderer2014, wachtel2015}; this significantly simplifies the calculations and enables to obtain analytical results even for relatively complex cases. The counting-field-dependent Liouvillian $\mathcal{W}(\lambda)$ can be obtained by inserting the appropriate counting fields into the Liouvillian $\mathcal{W}$~\cite{bagrets2003, bruderer2014}; its full matrix form is presented in the Appendix~\ref{sec:matrform}. The first $M$ scaled cumulants can be calculated by solving the following system of equations~\cite{bruderer2014, wachtel2015}:
\begin{align} \label{calccum}
\begin{cases}
\left \{ \frac{\partial^i}{\partial \lambda^i} \det[\chi(\lambda)-W(\lambda)] \right \}_{\lambda=0} =0, \\
\left[ \frac{\partial^i}{\partial \lambda^i} \chi(\lambda) \right]_{\lambda=0} = c_i, \\
\chi(0)  =  0,
\end{cases}
\end{align}
for $i=1, \dots , M$. For the considered case all cumulants of energy and heat are simple functions of cumulants of the particle current. A more general way to calculate the full counting statistics of energy and heat within the master equation approach can be found in the paper of S\'{a}nchez and B\"{u}ttiker~\cite{sanchez2012}.

Dynamics of the single-level system can be easily described using the master equation; cf. Ref.~\cite{bagrets2003} for details. Also in this case the results are equivalent to the obtained using the Levitov-Lesovik formula in the weak coupling limit.

\subsection{\label{sec:resultsint}Results}
Now I analyze the results, focusing on the strong Coulomb blockade regime ($U \rightarrow \infty$), for which the double occupancy of the system is forbidden; the results are compared with the noninteracting case $(U=0)$.  As in the previous section, symmetric coupling to the leads ($\Gamma_L=\Gamma_R=\Gamma$) and equal renormalized orbital energies ($\tilde{\epsilon}_L=\tilde{\epsilon}_R=0$) are assumed. 

%
\begin{figure} 
	\centering
	\includegraphics[width=0.9\linewidth]{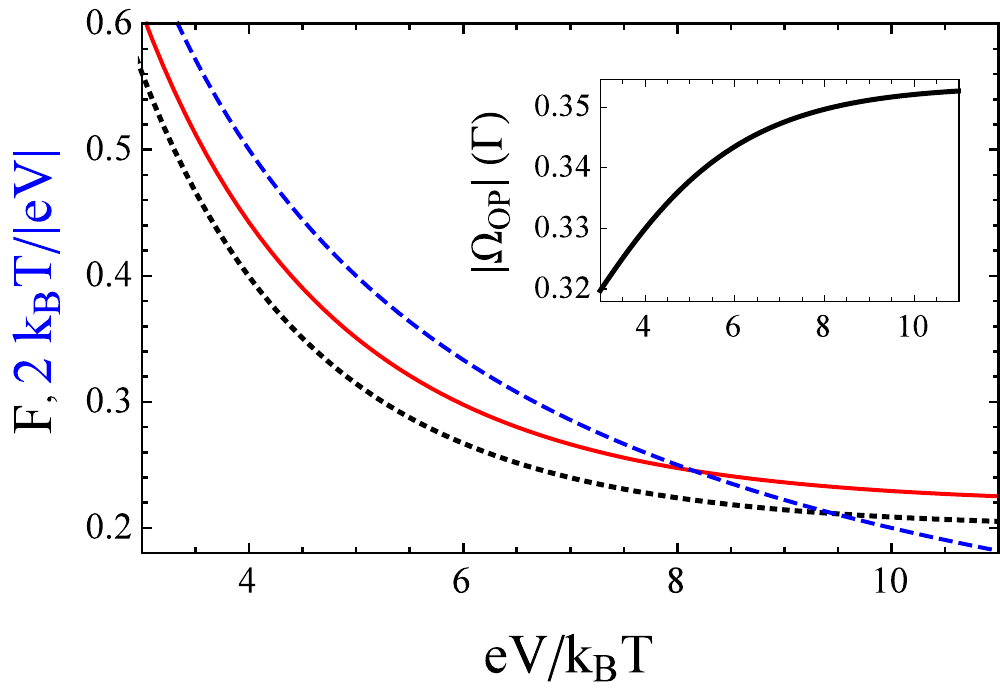} 
	\caption{Fano factor $F$ for the two-level bridge in the noninteracting case (red solid line) and the interacting case (black dotted line) as a function of the voltage $V$ compared with the classical bound $2k_B T/|eV|$ [cf. Eq.~\eqref{boundsim}, blue dashed line] for $\mu_L=-\mu_R=eV/2$ and $T_L=T_R=T$. Results for the noninteracting case obtained using the Levitov-Lesovik formula with $\Gamma=0.01 k_B T$ and $|\Omega|=\sqrt{15}\Gamma/6$. Results for the interacting case obtained using the master equation for the optimal value of $|\Omega|=|\Omega_{OP}|$, which dependence on $eV/k_B T$ is plotted in the top-right corner of the figure.}
	\label{fig:fanoint}
\end{figure}
%
Let us first find the conditions, for which the Fano factor is minimized. As previously mentioned, in the noninteracting case it takes place for $|\Omega|=\sqrt{15}\Gamma/6 \approx 0.65 \Gamma$. In contrast, in the interacting case the optimal value of $|\Omega|$, denoted further as $|\Omega_{OP}|$, depends on $\Gamma$, $f_L$ and $f_R$ in a non-trivial way, but does not exceed $\sqrt{3}\Gamma/2 \approx 0.87 \Gamma$. Figure~\ref{fig:fanoint} shows the Fano factor for the optimal value of $|\Omega|$ for the noninteracting and the interacting case compared with the classical limit. As one can clearly see, when the Coulomb interaction is present the current fluctuations are reduced more significantly than in the noninteracting system. This can be qualitatively explained as follows: Let us assume, that the system is initially in the state $|R \rangle$. In the noninteracting case, there are three possible trajectories of the subsequent evolution: (a)~coherent transition $|R \rangle \rightarrow |L \rangle$, (b)~transition $|R \rangle \rightarrow |0 \rangle$ due to electron tunneling from the right orbital to the right lead, (c)~transition $|R \rangle \rightarrow |D \rangle$ due to electron tunneling from the left lead to the left orbital. In the interacting case, due to the Coulomb blockade, the option (c)~is excluded. Therefore, since fewer trajectories of the evolution of the system are possible, the dynamics is less random, and therefore the current fluctuations are suppressed.

%
\begin{figure} 
	\centering
	\includegraphics[width=0.95\linewidth]{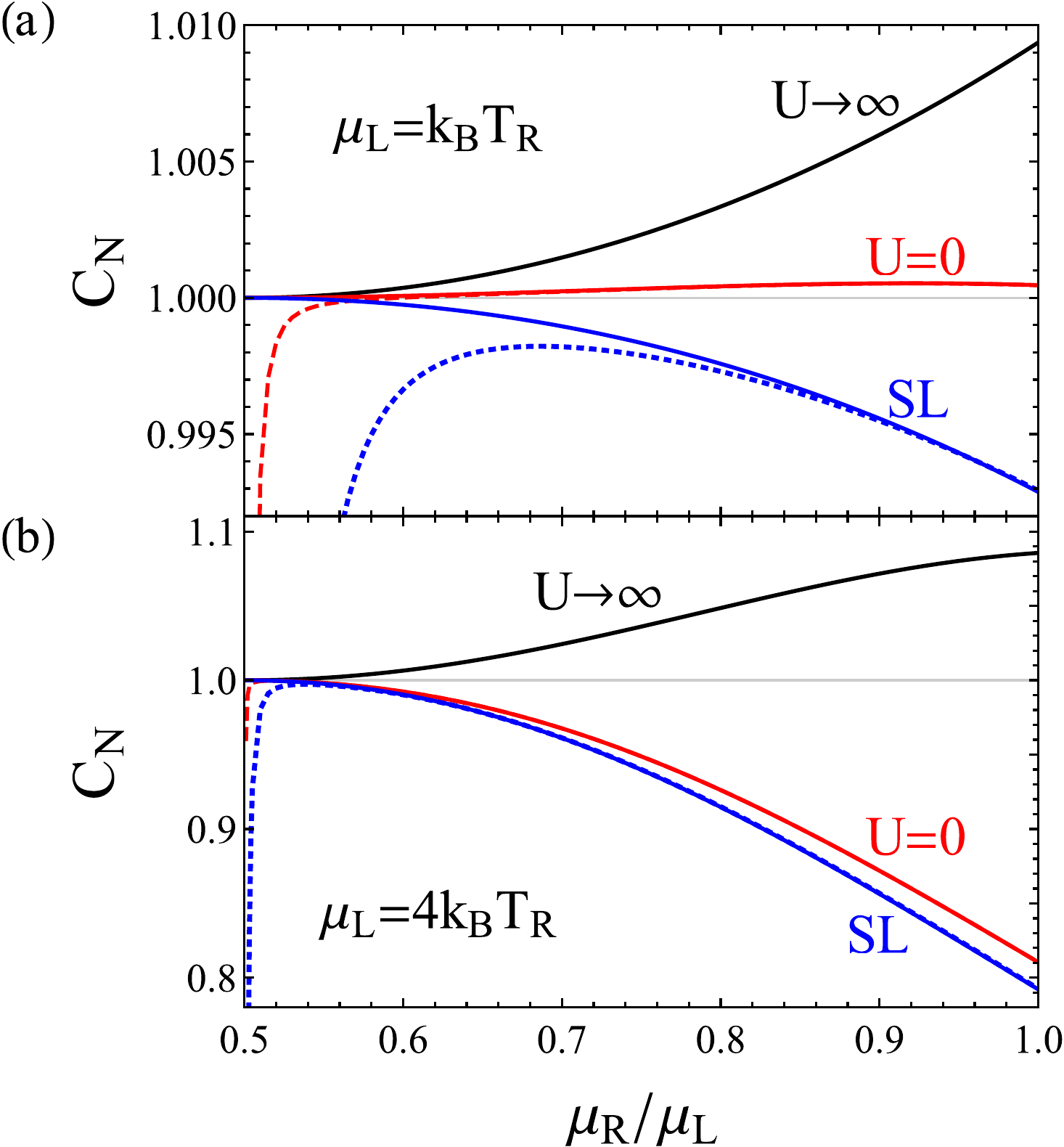} 
	\caption{Normalized constancy $C_N$ as a function of $\mu_R/\mu_L$ compared to the classical bound $C_N=1$, calculated by means of the master equation for the interacting case (black solid line, captioned as $U \rightarrow \infty$), the noninteracting case (red solid line, captioned as $U=0$) and the single-level system (blue solid line, captioned as ``SL''), for $T_L=2 T_R$, the optimal value of $|\Omega|$ (not shown) and $\mu_L=k_B T_R$~(a) or $\mu_L=4k_B T_R$~(b). The results are compared to the obtained using the Levitov-Lesovik formula with $\Gamma=0.002 k_B T_R$ for the noninteracting case (red dashed line) and the single-level system (blue dotted line); in the panel~(b) the red solid line and the red dashed line overlap in nearly whole range of $\mu_R/\mu_L$.}
	\label{fig:constint}
\end{figure}
%
Next, I analyze the normalized constancy $C_N$ [cf. Eqs.~\eqref{const}--\eqref{constsysweak}] as a function of the ratio $\mu_R/\mu_L$, focusing on the case of a reasonably small temperature bias $T_L=2T_R$. For the master equations results the weak-coupling formula for the constancy is applied [Eq.~\eqref{constsysweak}]. As shown in Fig.~\ref{fig:constint}~(a), for relatively small $\mu_L=k_B T_R$ the normalized constancy of the two-level bridge can exceed the classical limit (i.e., $C_N$ can be above 1) in both the interacting and the noninteracting case, but the improvement is not very pronounced (below 1\% in the interacting case and 0.1\% in the noninteracting case). On the other hand, for higher $\mu_L=4k_B T_R$ the normalized constancy is significantly enhanced in the interacting case -- it exceeds the classical limit by about 8\% for $\mu_R/\mu_L \rightarrow 1$ [Fig.~\ref{fig:constint}~(b)]; in the noninteracting case it drops, however, below the classical bound. 

In the presented cases, the improvement of constancy is particularly notable for $\mu_R/\mu_L$ close to 1 (for higher $\mu_L \gtrapprox 4.5 k_B T_R$ the maximum of $C_N$ is shifted to smaller values of $\mu_R/\mu_L$). In such a regime master equation results for the noninteracting case and the single-level system well agree with the exact results obtained for small $\Gamma=0.002 k_B T_R$; in particular, for $\mu_L=4k_B T_R$ the results converge in nearly whole range of $\mu_R/\mu_L$ [Fig.~\ref{fig:constint}~(b)]. This confirms the applicability of the master equation for $\mu_R/\mu_L$ sufficiently higher than $T_R/T_L$, which supports the claim, that results for the interacting case in this parameter regime are also physically relevant. As in Fig.~\ref{fig:normconstnon}, the master equations predictions and the exact results deviate for $\mu_R/\mu_L \approx T_R/T_L$, for which the Fermi distributions of the left and the right lead become comparable in the vicinity of the orbital levels (i.e., the thermal bias becomes counter-balanced by the voltage bias, which leads to the current suppression); this illustrates the limits of applicability of the master equation approach.

%
\begin{figure}
	\centering
	\includegraphics[width=0.95\linewidth]{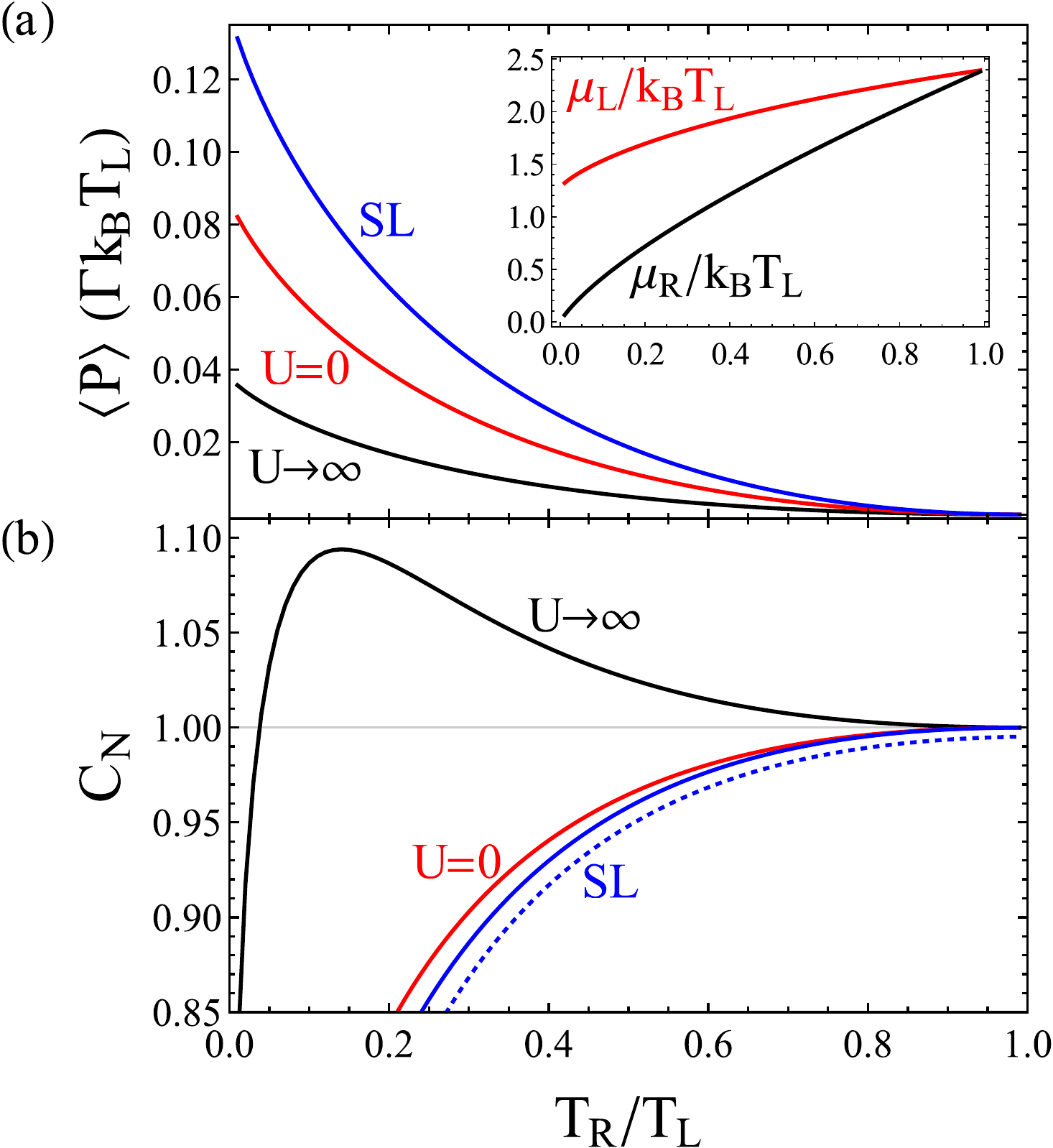} 
	\caption{Mean power $\langle P \rangle$~(a) and the normalized constancy $C_N$~(b) as a function of $T_R/T_L$ calculated by means of the master equation for the interacting case (black solid line, captioned as $U \rightarrow \infty$), the noninteracting case (red solid line, captioned as $U=0$) and the single-level system (blue solid line, captioned as ``SL''), for electrochemical potentials as plotted in the top-right corner of the panel~(a) and the optimal value of $|\Omega|$ (not shown). Blue dotted line in the panel~(b) shows the normalized constancy for the single-level system calculated using the Levitov-Lesovik formula with $\Gamma=0.01 k_B T_L$; the analogous plot for the noninteracting two-level bridge overlaps with the red solid line.}
	\label{fig:powconst}
\end{figure}
%
As mentioned, for sufficiently small $\mu_L$ the normalized constancy in the interacting case is optimized for $\mu_R/\mu_L \rightarrow 1$. In this limit the ratio of the power variance to the mean power $\text{Var}(P)/\langle P \rangle=F(\mu_R-\mu_L)$ tends to 0. However, in such a limit the mean power $\langle P \rangle=\langle I \rangle(\mu_R-\mu_L)$ also drops to 0. From a practical point of view it is more relevant to analyze the case of a finite power output. Usually one studies the performance at maximum power; however, in the considered system the power is maximized for $|\Omega| \rightarrow \infty$, for which fluctuations are not reduced below the classical bound. To enable some meaningful analysis, let us compare the performance of thermoelectric generators based on the two-level bridge and the single-level system as a function of $T_R/T_L$, with parameters chosen in the following way: (a) first, the electrochemical potentials, for which the power of the single-level generator is maximized, are applied; (b) secondly, power fluctuations of the two-level bridge are minimized over $|\Omega|$.

Since in the weak coupling regime efficiency equals $\eta=1-\mu_R/\mu_L$, it is the same in all cases. On the other hand, power of the two-level bridge generator is significantly lower than of the single-level system. This is due to the fact that tunneling between the orbitals requires some additional time. The reduction is more significant in the interacting case due to the Coulomb blockade, which forbids tunneling to the system when it is singly occupied. However, since power fluctuations are suppressed more strongly than the mean power, the normalized constancy of the two-level bridge is increased in comparison with the single-level system. A significant enhancement of constancy (above the classical limit) is observed, however, only in the interacting case. The non-monotonous character of dependence of $C_N$ on $T_R/T_L$ for the interacting two-level bridge has no obvious physical interpretation; it seems to be a consequence of the procedure used to define values of the system parameters. It does not indicate the presence of any noise enhancement mechanism for low $T_R/T_L$, since $C_N$ can be still above 1 for other values of electrochemical potentials (not shown). Quite surprisingly, whereas one can observe a small but non-negligible difference of the master equation results and those obtained using the Levitov-Lesovik formula for the single-level system with moderately weak coupling to the leads $\Gamma=0.01 k_B T_L$ [Fig.~\ref{fig:powconst}~(b), blue solid line and blue dotted line, respectively], hardly any deviation is observed for the noninteracting two-level bridge (corresponding plots overlap); it becomes significant only for stronger tunnel coupling $\Gamma \gtrapprox 0.05 k_B T_L$ (not shown). This is another illustration of applicability of the master equation approach in the weak coupling regime.

\section{\label{sec:conclusions}Conclusions}
The paper analyzes fluctuations of the output power of the thermoelectric generator based on the two-level bridge, i.e. two tunnel-coupled orbitals, each attached to a separate electronic reservoir. Transport is studied by means of the exact Levitov-Lesovik formula in the situation, when the Coulomb interaction between orbitals is neglected, as well as using the quantum master equation in the interacting case. In comparison with the thermoelectric generator based on a single electronic level, in the analyzed system transport of the electron between the leads involves an additional intermediate process, namely the coherent tunneling of electrons between the orbitals. This process is associated with a unitary evolution of the electron state instead of a stochastic Poisson transition corresponding to the sequential tunneling. Due to the partially unitary (instead of purely stochastic) character of the evolution of the system the current and power fluctuations are suppressed, which improves constancy of the heat engine. This effect is particularly significant when the Coulomb interaction between the electrons is present; this is because the Coulomb blockade reduces the set of possible stochastic trajectories of the evolution of the system, which decreases randomness of the dynamics. 

Most significantly, the reduction of power fluctuations enables to overcome the thermodynamic trade-off between efficiency, power and constancy which applies to classical, purely stochastic Markovian steady state heat engines~\cite{pietzonka2017}. The study shows, therefore, an example of the positive influence of the quantum coherent unitary evolution on performance of heat engines. The presented results may motivate the search for similar constancy enhancement in other types of quantum heat engines, e.g., optical~\cite{harbola2012, scully2010, scully2011} or superconducting~\cite{hardal2017, karimi2016} ones, as well as in multi-terminal ballistic junctions~\cite{brandner2017} or classical systems with unitary Hamiltonian component of the dynamics~\cite{fischer2018}.

\section*{Acknowledgments}
I thank B. R. Bu\l{}ka for the careful reading of the manuscript and the valuable discussion. This work has been supported by the National Science Centre, Poland, under the Project No. 2016/21/B/ST3/02160. 

\appendix

\section{Hamiltonian of the quantum dot spin valve} \label{sec:hamsv}
Hamiltonian of a single-level spinfull quantum dot attached to the spin-polarized leads magnetized in the z-direction can be written as~\cite{rudzinski2001, braun2004}
\begin{align} \label{hamsv}
& \hat{H}_{sv} =  \sum_{\sigma} \epsilon_{\sigma} d_\sigma^\dagger d_\sigma +\gamma \mathbf{B} \cdot \hat{\mathbf{s}}+U d^\dagger_\uparrow d_\uparrow d^\dagger_\downarrow d_\downarrow \\ \nonumber
&+\sum_{\alpha \sigma \mathbf{k}} \epsilon_{\alpha \sigma \mathbf{k}} c_{\alpha \sigma \mathbf{k}}^{\dagger} c_{\alpha \sigma \mathbf{k}}+\sum_{\alpha \sigma \mathbf{k}} \left(t_{\alpha \sigma} c_{\alpha \sigma \mathbf{k}}^{\dagger} d_{\sigma} +t^*_{\alpha \sigma} d_{\sigma}^\dagger c_{\alpha \sigma \mathbf{k}} \right),
\end{align}
where $\sigma \in \{ \uparrow, \downarrow \}$ denotes the spin, $\gamma$ is the gyromagnetic ratio (taken to be isotropic), $\mathbf{B}$ is the magnetic field and $ \hat{\mathbf{s}}$ is the spin operator. Other symbols are analogous to the used in Eq.~\eqref{hamtotal}. The spin-dependent coupling strength to the lead $\alpha$ reads then $\Gamma_{\alpha}^\sigma = 2 \pi |t_{\alpha \sigma}|^2 \rho_{\alpha}^\sigma$, where $\rho_{\alpha}^\sigma$ is the density of states for the electrons with spin $\sigma$ in the lead $\alpha$. For the magnetic field directed along the x-axis one obtains $\mathbf{B} \cdot \hat{\mathbf{s}}=\hbar B \sigma^x/2={\hbar {B(c^\dagger_\uparrow c_\downarrow+c^\dagger_\downarrow c_\uparrow)}/2}$, where $\sigma^x$ is the Pauli-X matrix. In the case of perfect and antiparallel spin polarization of the leads ($\rho_L^\downarrow=\rho_R^\uparrow=0$) the Hamiltonian~\eqref{hamsv} becomes then mathematically equivalent to the Hamiltonian~\eqref{hamtotal}, with $\gamma \hbar B/2$ corresponding to $\Omega$. 

\section{Full matrix form of the Liouvillian} \label{sec:matrform}
The counting field dependent Liouvillian $\mathcal{W}(\lambda)$ for the two-level bridge reads	
\begin{widetext}
\begin{align} \label{liovchi} &\mathcal{W}(\lambda)=
\begin{pmatrix}
-\sum_\alpha \Gamma_\alpha f_\alpha & \Gamma_L (1-f_L) & \Gamma_R (1-f_R) e^\lambda & 0 & 0 & 0\\
\Gamma_L f_L & -\Gamma_L (1-f_L)-\Gamma_R f_R^U & 0 & \Gamma_R (1-f_R^U) e^\lambda & 0 &-2 \Omega \\
\Gamma_R f_R e^{-\lambda} & 0 & -\Gamma_L f_L^U-\Gamma_R (1-f_R) & \Gamma_L (1-f_L^U) & 0 & 2 \Omega\\
0 &  \Gamma_R f_R^U e^{-\lambda} & \Gamma_L f_L^U &-\sum_{\alpha} \Gamma_\alpha (1-f_\alpha^U) & 0 & 0 \\
0 & 0 & 0 & 0 & \Gamma_D & \Delta \tilde{\epsilon} \\
0 & \Omega & -\Omega & 0 & -\Delta \tilde{\epsilon} & \Gamma_D
\end{pmatrix},
\end{align}
\end{widetext}
with $\alpha \in \{L, R\}$, $f_\alpha$ and $f_\alpha^U$ defined below Eq.~\eqref{lindbladop}, $\Delta \tilde{\epsilon}=\tilde{\epsilon}_L-\tilde{\epsilon}_R$ and $\Gamma_D=\sum_{\alpha} \Gamma_\alpha (f_\alpha-f_\alpha^U-1)/2$. The Liovillian $\mathcal{W}$ is obtained by taking $\lambda=0$.

\end{document}